# Paraconductivity and magnetoconductivity approaches to superconducting fluctuations in La$_{1.6-x}$Nd$_{0.4}$Sr$_x$CuO$_4$ films

Y. Liu[*] and S. L. Wan

*Department of Modern Physics, University of Science and Technology of China, Hefei 230026, P. R. China*

X. G. Li

*Hefei National Laboratory for Physical Sciences at Microscale, Department of Physics, University of Science and Technology of China, Hefei 230026, P. R. China*

Abstract

The in-plane conductivity induced by superconducting fluctuations above the superconducting transition, i.e. the in-plane paraconductivity $\Delta\sigma_{ab}(T)$, was extracted from the low-temperature upturn region of the in-plane resistivity $\rho_{ab}(T)$ in the reduced-temperature range $10^{-2} \leq \varepsilon = \ln(T/T_c) \leq 1$ for the *c*-axis oriented La$_{1.6-x}$Nd$_{0.4}$Sr$_x$CuO$_4$ films (*x* = 0.1, 0.12, 0.14, 0.16). It is then shown that this $\Delta\sigma_{ab}(T)$ can be explained with the short-wavelength cutoff theory presented by Vidal *et al*. The obtained out-of-plane coherence length, $\xi_c(0)$, which varies from 0.9 to 1.8 Å with increasing Sr content *x*, shows the similar Sr doping dependence as observed in La$_{2-x}$Sr$_x$CuO$_4$ films by a traditional Aslamazov-Larkin theory. Our results suggest, therefore, that Nd doping mainly affects the normal-state properties in the samples, and has little influence on the superconducting fluctuations. The resistive transition broadening of La$_{1.6-x}$Nd$_{0.4}$Sr$_x$CuO$_4$ films (*x* = 0.1, 0.12, 0.14, 0.16) was measured under magnetic fields up to 14 T oriented both parallel and perpendicular to the *c* axis. The resistive transition displays characteristic broadening behavior, depending on Sr content *x*, for $H \parallel c$. For *x* = 0.1 superconducting transition curves shows the so-called fan-shaped broadening, while for *x* exceeding 0.12, the application of a field simply causes a parallel shift of the curves to lower temperatures. The field-induced fluctuation conductivity in La$_{1.6-x}$Nd$_{0.4}$Sr$_x$CuO$_4$ films, i.e. the in-plane magnetoconductivity $\Delta\sigma_{ab}(H,T)$, was extracted in temperatures up to 1.8$T_c$. We find that the $\Delta\sigma_{ab}(H,T)$ data in the case of $H \parallel c$ can be well described as fluctuation conductivity but only with the Aslamasov-Larkin orbital term, in which the in-plane coherence length $\xi_{ab}(0)$ is free parameter. The anisotropy ratio $\xi_{ab}(0)/\xi_c(0)$ estimated from these results reduces continuously with increasing Sr content *x*. We suggest that the transformation from the fan-shaped broadening to parallel shift in La$_{1.6-x}$Nd$_{0.4}$Sr$_x$CuO$_4$ films may be interpreted in terms of a dimension crossover from 2D to 3D with increasing charge carrier density.

---

[*] Corresponding author. Email address: yliu@ustc.edu.cn

PACS number(s): 74.25.Fy, 74.72.Dn, 74.40.+k

## I. INTRODUCTION

Owing to strong anisotropy, high critical temperature, and low charge carrier density, thermodynamic fluctuations are greatly enhanced in high $T_c$ superconductors (HTS). Thermal fluctuations result in a finite probability of a Cooper pair formation above the critical temperature $T_c$, leading to an excess conductivity $\Delta\sigma$ [1-3], which is one of the interesting aspects of HTS [4]. Experimentally, the resistivity $\rho(T)$ of HTS presents a pronounced rounding of the superconducting transition due to superconducting fluctuations [5]. A well documented peak in $c$-axis resistivity $\rho_c(T)$ in $Bi_2Sr_2CaCu_2O_{8+x}$ (BSCCO) just above $T_c$ can be explained within the framework of superconducting fluctuations [6]. Moreover, fluctuation effects in HTS may persist up to quite high temperatures. For underdoped HTS in-plane resistivity $\rho_{ab}(T)$ exhibits a downward bending behavior and deviates from its high-temperature linear-$T$ dependence below the pseudogap temperature $T^*$ [7]. Although the possibility of fluctuation pairing in HTS at temperatures much higher than $T_c$ is widely discussed in the literature [7,8], it is still under debate whether superconducting fluctuations are linked to the opening of normal-state pseudogap in underdoped HTS [9,10].

Gained a number of experimental evidences, the charge stripe theory has thrown light on the origin of the pseudogap and the mechanism responsible for the high $T_c$ superconductivity [11,12]. As we know, any movement of introduced charges tends to create strings of misaligned spins in the antiferromagnetic (AF) aligned $CuO_2$ planes, but the strings of broken AF bonds can be erased by another charge moving in the same direction. One may suspect that one moving charge might distort nearby spins in such a way as to exert an attractive force on other charges, binding them together into pairs [13]. Doped charges may aggregate into stripes at high temperatures due to the large exchange energy $J$ [14]. Strong evidence comes from that in the lightly doped single crystal of $La_{2-x}Sr_xCuO_4$ (LSCO), $\rho_{ab}(T)$ is insensitive to the long-range magnetic order, and keeps its metallic behavior well below the Néel temperature $T_N$ [15]. The picture is quite appealing: Doped holes move along the stripe without disturbing the underlying AF $CuO_2$ planes. Now that the Cooper pairs form far above $T_c$, the investigation of superconducting fluctuations is useful to reveal the nature of pairing in the charge stripe scenario.

Considerable attention has been given to the $La_{1.6-x}Nd_{0.4}Sr_xCuO_4$ system due to the coexistence of, and competition between, superconductivity and charge-stripe order in the materials [16,17]. The purpose of the present work is to investigate the relation among superconducting fluctuations, normal-state transport properties and the charge stripe order. Samples used in this work are single-crystal-like $La_{1.6-x}Nd_{0.4}Sr_xCuO_4$ films ($x$ = 0.1, 0.12, 0.14, 0.16) grown by an off-axis magnetron sputtering technique [18]. The temperature and magnetic field dependences of $\rho_{ab}(T)$ for $La_{1.6-x}Nd_{0.4}Sr_xCuO_4$ films were measured in an Oxford 15 T superconducting magnet for both parallel and perpendicular to the $c$ axis. In $La_{1.6-x}Nd_{0.4}Sr_xCuO_4$ films, the static charge stripes observed in the single crystals become dynamically fluctuating charge stripes due to epitaxial strain between the films and substrates [18]. Upon cooling, a low-temperature upturn of $\rho_{ab}(T)$ curves is observed in optimally doped $La_{1.6-x}Nd_{0.4}Sr_xCuO_4$ films, which discriminates the Nd doped LSCO films from Nd free samples. The low-temperature upturn behavior in $La_{1.6-x}Nd_{0.4}Sr_xCuO_4$ films can be attributed to a lattice randomness effect by Nd substitution. With the suppression of static charge stripe order, there is a possibility that charge stripes tend to meander through $CuO_2$ planes, which is irrelevant to the physics of charge stripes by analogy with the case of bulk crystals [19-21].

**II. ANALYSIS ON PARACONDUCTIVITY**

One of the magnitudes best adapted, from both the theoretical and the experimental point of view, to study superconducting fluctuations in HTS is so-called fluctuation conductivity [1-4]. When a magnetic field is absent, fluctuation conductivity, usually referred to as "paraconductivity", is defined as

$$\Delta\sigma_{ab}(T) \approx \frac{1}{\rho_{ab}(T)} - \frac{1}{\rho_{abB}(T)}, \qquad (1)$$

where $\rho_{ab}(T)$ is the measured in-plane resistivity, and $\rho_{abB}(T)$ is the extrapolated normal-state resistivity without the fluctuation effects (suffix $B$ means background).

Generally, the interpretation of experimental data in terms of fluctuation theory requires the extrapolation of the normal-state property from high temperatures, where fluctuations effects are supposed to be negligible. Therefore, the investigations of paraconductivity in HTS rely crucially on an accurate estimate of the normal state properties. Figure 1 shows the superconducting transition curves of $La_{1.6-x}Nd_{0.4}Sr_xCuO_4$ films ($x$ = 0.1, 0.12, 0.14, 0.16). The mean-field critical temperature $T_c$ of the films is determined from the peak value in the plot of $d\rho_{ab}/dT$ vs $T$. A low-temperature resistivity upturn is observed in the optimally doped sample $x = 0.16$, which discriminates the charge transport of $La_{1.6-x}Nd_{0.4}Sr_xCuO_4$ films from Nd-free LSCO films [22,23]. The increase of $\rho_{ab}(T)$ with decreasing temperature approximately follows the $\log(1/T)$ transport (see Ref. 18), as observed in $La_{1.48}Nd_{0.4}Sr_{0.12}CuO_4$ single crystal [17]. Here, Nd substitution gives rise to a lattice randomness effect, which resembles local pinning centers and slows down the transverse fluctuations of charge stripes as observed in lightly Zn doped $La_{2-x}Sr_xCu_{0.98}Zn_{0.02}O_4$ single crystals [24].

Near optimum doping a straight line fits the resistivity data very well. Thus the formula $\rho_{ab}(T) = a + bT$ is then extrapolated to low temperature and used to extract the fluctuation contribution to the conductivity. In the case of underdoped samples in the pseudogap region, however, an upward curvature of $\rho_{ab}(T)$ appears at low temperature above $T_c$. Obviously, the extrapolation with the linear temperature dependence of background resistivity becomes invalid. A pioneer treatment for upturn resistivity above $T_c$ was presented by Currás *et al*. [10]. A polynomial $a_1/T + a_2 + a_3T + a_4T^2$ was suggested to extract paraconductivity in the underdoped LSCO films. Obviously, the fittings to the $\rho_{ab}(T)$ data with this polynomial become crucial for the correct description of the paraconductivity. If one assumes the normal state to follow closely the resistivity data around $T_c$, there is little space for the paraconductivity contribution, which is rapidly suppressed above $T_c$, and vice versa. In this study, the fitting procedure is similar to the one already used in Ref. 10. Here, the fitting were done in the temperature region $3T_c \leq T \leq 8T_c$. We also require that $\rho_{abB}(T)$ does not produce a negative paraconductivity at any temperature and fitting parameters $a_1, a_2, a_3, a_4$ always keep positive. The results are shown by the dashed lines in Fig. 1.

Figure 2 shows the plots of extracted in-plane paraconductivity $\Delta\sigma_{ab}(T)$ in $La_{1.6-x}Nd_{0.4}Sr_xCuO_4$ films. The magnitude in paraconductivity amplitude of $La_{1.6-x}Nd_{0.4}Sr_xCuO_4$ films not only is quite close to that extracted from $La_{2-x}Sr_xCuO_4$ films at the same Sr doping level, but also shows the similar temperature dependence [10]. For HTS, the fluctuation conductivity has been most extensively studied within the framework of Aslamazov-Larkin (AL) theory [1,2]. As the temperatures are very close to $T_c$, a 3D AL expression holds, and

$$\Delta\sigma_{AL}^{3D}(\varepsilon) = \frac{e^2}{32\hbar\xi_c(0)} \frac{1}{\varepsilon^{1/2}}, \qquad (2)$$

while for temperatures even slightly above $T_c$ a 2D AL formulation is appropriate:

$$\Delta\sigma_{AL}^{2D}(\varepsilon) = \frac{e^2}{16\hbar d}\frac{1}{\varepsilon}. \qquad (3)$$

Here $\varepsilon = \ln(T/T_c)$ is the reduced temperature, $d$ is the spacing of the $CuO_2$ planes (~6.5 Å in $La_{1.6-x}Nd_{0.4}Sr_xCuO_4$ films), and $\xi_c(0)$ is the out-of-plane coherence length at the absolute zero temperature. These expressions account well for the fluctuating regime near $T_c$ both in optimally doped and underdoped HTS.

The dashed lines in Fig. 2 are the plots of a 2D AL behavior. Remarkably, the 2D AL theory of paraconductivity does not allow for any fitting parameters besides the experimentally well-accessible distance $d$. It is found that the $\Delta\sigma_{ab}(\varepsilon)$ data of $La_{1.6-x}Nd_{0.4}Sr_xCuO_4$ films approximately reproduce the slope of -1 as predicted by Eq. (3) in the moderate-$\varepsilon$ range $10^{-2} \leq \varepsilon \leq 1$, but the discrepancy between $\Delta\sigma_{ab}(\varepsilon)$ and the 2D AL theory gradually enlarges with increasing Sr content $x$. However, all the $\Delta\sigma_{ab}(T)$ curves of $La_{1.6-x}Nd_{0.4}Sr_xCuO_4$ films bend down in the high-$\varepsilon$ region according to the $1/\varepsilon^3$ law, as shown by the dotted lines in Fig. 2. The paraconductivity falls well below predicted theoretical values in the high temperature limit. This is attributed to the reduced role of high-wave-vector contributions to the paraconductivity while the short-wavelength fluctuations become important [25]. In order to reconcile the discrepancy between the theoretical $\Delta\sigma(\varepsilon)$ and the experimental results in the high-$\varepsilon$ region, a phenomenological short-wavelength cutoff should be introduced in the theoretical fluctuation spectrum [26]. As a further attempt to understand the behavior of the thermal fluctuations of Cooper pairs in the high-$\varepsilon$ region ($\varepsilon \geq 0.1$) in HTS, Vidal *et al.* suggested that $\Delta\sigma_{ab}(\varepsilon)$ may be analyzed in terms of the mean-field-like Gaussian-Ginzburg-Landau (GGL) approach by means of the introduction of a total-energy cutoff, which takes into account both the kinetic energy and the quantum localization energy of each fluctuating mode [10,27]. Thus the analysis on paraconductivity can be well extended to the high-$\varepsilon$ region by the following expression:

$$\Delta\sigma_{ab}(\varepsilon)_E = \frac{e^2}{16\hbar d}\left[\frac{1}{\varepsilon}\left(1+\frac{B_{LD}}{\varepsilon}\right)^{-1/2} - \frac{2}{c} + \frac{\varepsilon+B_{LD}/2}{c^2}\right], \qquad (3)$$

where $B_{LD} = (2\xi_c(0)/d)^2$, $c$ is a constant close to 1. It should be pointed out that the 2D AL formulation can be obtained by applying condition $B_{LD} \ll \varepsilon$ in Eq. (3).

It is found that the $\Delta\sigma_{ab}(\varepsilon)$ data of La$_{1.6-x}$Nd$_{0.4}$Sr$_x$CuO$_4$ films are well described using Eq. (3), as shown by the solid lines in Fig. 2. The fittings were done in the $\varepsilon$ range between $10^{-2} \leq \varepsilon \leq 1$ with $\xi_c(0)$ and $c$ as free parameters, which cover nearly three orders of magnitude in paraconductivity amplitude. The obtained out-of-plane coherence length, $\xi_c(0)$, varies from 0.9 to 1.8 Å with increasing Sr content $x$. The Sr doping dependence of $\xi_c(0)$ in La$_{1.6-x}$Nd$_{0.4}$Sr$_x$CuO$_4$ films is similar to that observed in LSCO films by a traditional AL theory [28,29]. Our results strongly suggest, therefore, that the superconducting fluctuations in the samples that undergo a metal-insulator crossover above the superconducting transition can still be described. Furthermore, it implies that the freezing of charges into the stripes that is suggested to explain the low-temperature resistivity upturn in La$_{2-x-y}$Nd$_y$Sr$_x$CuO$_4$ and Zn doped LSCO systems [24] has little influence on the superconducting fluctuations, while the doping mainly affects the normal-state properties in the samples.

An interesting question arises from the verification of the stripe theory by the fluctuation conductivity approaches. Based on the charge stripe model [11,12], a theoretical sketch of the phase diagram of HTS may be that the holes aggregate into stripes below $T^*$, while the self-organized stripe arrays allow local AF correlations to develop in the hole-free regions of the sample. The global superconductivity of HTS is controlled by the Josephson coupling through the pair hopping between dynamically fluctuating stripes, which is required to establish phase coherence for an array of stripes. It should be made clear that static stripes are certainly bad for the Josephson coupling between stripes, whereas fluctuating stripes produce better Josephson coupling. Rather naturally, one may suppose that superconducting fluctuations become weak with the stabilization of charge stripes. Thus, the temperature range dominated by superconducting fluctuations tends to shrink while the pseudogap formation temperature $T^*$ decreases [30]. Even in the optimally doped sample $x = 0.16$, however, $T^*$ is quite close to the room temperature. Our results confirm that superconducting fluctuations are only established in the vicinity of $T_c$ in La$_{1.6-x}$Nd$_{0.4}$Sr$_x$CuO$_4$ films while the paraconductivity can not be measured with temperature approaching to $2T_c$. Therefore, the in-plane paraconductivity is independent of the opening of a pseudogap in the normal state of La$_{1.6-x}$Nd$_{0.4}$Sr$_x$CuO$_4$ films, as observed in LSCO films [10]. It is believed that these results are fully available in the case of single crystals due to the similar upturn behavior of $\rho_{ab}(T)$ at low temperatures. However, this by no means implies that preformed pairs are not present below $T^*$. It is suggested that the establishment of superconducting phase coherence in HTS is not due to a simple condensation of preformed pairs. In a cautious manner, it simply means that the superconducting coherence is driven by the formation of more loosely bound traditional BCS pairs [9,31]. Superconducting fluctuations become observable only when HTS achieve the phase coherence depending on the Josephson coupling between stripes, although Cooper pairs might form in the metallic conducting stripes at high temperatures.

**III. ANALYSIS ON MAGNETOCONDUCTIVITY**

The extrapolation of the normal-state properties of HTS is somewhat arbitrary when considering the correct choice of background resistivity. In the study of magnetoconductivity one can avoid this problem, where the field-independent normal-state contribution is canceled [32]. The magnetoconductivity is expressed as

$$\Delta\sigma_{ab}(H,T) = \frac{1}{\rho_{ab}(H,T)} - \frac{1}{\rho_{ab}(0,T)}. \tag{4}$$

Figure 3 shows the resistive transition curves of $La_{1.6-x}Nd_{0.4}Sr_xCuO_4$ films ($x = 0.1$, 0.12, 0.14, 0.16) under various fields for both $H \perp c$ and $H \parallel c$ configurations. Being of an intrinsic layered structure, $La_{1.6-x}Nd_{0.4}Sr_xCuO_4$ films exhibit different broadening behavior for $H \parallel c$ and $H \perp c$ configurations. As most HTS under $H \parallel c$ orientation, the resistive transition curves of $La_{1.6-x}Nd_{0.4}Sr_xCuO_4$ films display remarkably broadening. It is found that for low doping level $x = 0.1$, superconducting transition curves dramatically broaden and fan out with increasing magnetic fields, i.e., the onset temperature of superconducting transition keeps steady, while for the samples $x = 0.12, 0.14, 0.16$, the transition curves shift to lower temperatures almost in parallel, similar to those observed in $La_{1.6-x}Nd_{0.4}Sr_xCuO_4$ single crystals ($x = 0.12, 0.15$) [33,34] and LSCO single crystal ($x = 0.12$) [35].

Generally, for conventional superconductors, the superconducting transition curve in magnetic fields shifts to the low-temperature side in parallel with increasing field, which is attributed to the small superconducting fluctuation originating from the large superconducting coherence length. For HTS, characteristic of intrinsic layered structure, the superconducting transition curve shows the fan-shape broadening behavior in the underdoped regime, which results from the large superconducting fluctuation due to the small superconducting coherence length [35].

To get further insight into this feature, we estimate the contribution from the superconducting fluctuations to elucidate the crossover from the HTS-like behavior to the conventional-superconductor-like behavior in $La_{1.6-x}Nd_{0.4}Sr_xCuO_4$ films. The present analysis method of the magnetoconductivity in HTS was originally treated by Hikami and Larkin (HL) [36], Maki and Thompson (MT) [37], and Aronov *et al*. [38]. In layered HTS, the orbital effect is confined within the $CuO_2$ planes for $H \parallel c$, while the Zeeman effect is independent of the field direction. The magnetoconductivity for $H \parallel c$ comprises four contributions: the AL-orbital (ALO), MT-orbital (MTO), AL-Zeeman (ALZ), and MT-Zeeman (MTZ) terms

$$H \parallel c : \Delta\sigma(H,\varepsilon) = \Delta\sigma_{ALO} + \Delta\sigma_{MTO} + \Delta\sigma_{ALZ} + \Delta\sigma_{MTZ}. \tag{5}$$

For $H \perp c$, only the Zeeman terms contribute to the magnetoconductivity,

$$H \perp c : \Delta\sigma(H,\varepsilon) = \Delta\sigma_{ALZ} + \Delta\sigma_{MTZ}. \tag{6}$$

Each term of Eqs. (5) and (6) is expressed as [39],

$$\Delta\sigma_{ALO} = \frac{e^2}{8\hbar} \int_0^{2\pi/d} \frac{1}{\varepsilon_k} \left(\frac{\varepsilon_k}{b}\right)^2 \left[\psi\left(\frac{1}{2}+\frac{\varepsilon_k}{2b}\right) - \psi\left(1+\frac{\varepsilon_k}{2b}\right) + \frac{b}{\varepsilon_k}\right]\frac{dk}{2\pi} - \frac{e^2}{16\hbar d}\frac{1}{\varepsilon\sqrt{1+2\alpha}}, \tag{7}$$

$$\Delta\sigma_{MTO} = \frac{e^2}{8\hbar} \frac{1}{d\varepsilon(1-\alpha/\delta)} \left\{ \int_0^{2\pi} \left[\psi\left(\frac{1}{2}+U\right) - \psi\left(\frac{1}{2}+V\right)\right]\frac{dx}{2\pi} - \ln\left(\frac{\delta}{\alpha}\frac{1+\alpha+\sqrt{1+2\alpha}}{1+\delta+\sqrt{1+2\delta}}\right)\right\}, \tag{8}$$

$$\Delta\sigma_{ALZ} = -7\zeta(3)\frac{e^2}{16\hbar d\varepsilon^2}\frac{1+\alpha}{(1+\alpha)^{3/2}}\left(\frac{g\mu_B B}{4\pi k_B T_c(0)}\right)^2, \tag{9}$$

$$\Delta\sigma_{MTZ} = -7\zeta(3)\frac{e^2}{8\hbar d\varepsilon^2}\frac{1}{1-\alpha/\delta}\left(\frac{g\mu_B B}{4\pi k_B T_c(0)}\right)^2 \times \left[\frac{1}{1-\alpha/\delta}\ln\left(\frac{\delta}{\alpha}\frac{1+\alpha+\sqrt{1+2\alpha}}{1+\delta+\sqrt{1+2\delta}}\right) - \frac{1}{\sqrt{1+2\alpha}}\right] \tag{10}$$

where $\alpha = (2\xi_c^2(0))/(d^2\varepsilon)$, $\delta = (16\xi_c^2(0)k_B T\tau_\phi)/(\pi\hbar d^2)$, $\varepsilon_k = \varepsilon[1+\alpha(1-\cos kd)]$, $b = 2e\xi_{ab}^2(0)H/\hbar$, $U = \varepsilon[1+\alpha(1-\cos x)]/2b$, $V = (\alpha\varepsilon/2b)(1/\delta+1-\cos x)$, $\Psi$ is the digamma function, $\zeta$ is Riemanns zeta function, and $g$ is the gyromagnetic ratio (~2), $\xi_{ab}(0)$ is the in-plane coherent length at the absolute zero temperature, and $\tau_\phi$ is the phase relaxation time. These expressions are derived on the basis of the Ginzburg-Landau theory applied to layered superconductors in the dirty limit.

In Fig. 4, the magnetoconductivity data of the samples $x = 0.1$ and 0.16 for both $H \parallel c$ (open circles) and $H \perp c$ (open squares) configurations are plotted as a function of $H$. In order to avoid the interference of the flux motion, the magnetoconductivity data for both magnetic orientations were extracted in temperatures up to $1.8T_c$, which is ~10 K above $T_c$. For $H \parallel c$, solid line represent the sum of ALO, MTO, ALZ, and MTZ terms, resulting from Eqs. (7)-(10). It is clearly seen that the magnetoconductivity for $H \parallel c$ is well described with the ALO term, while the MTO and MTZ terms are almost negligible for both $x = 0.1$ and 0.16 samples. For $H \perp c$, there are considerable deviations between the measured data and the ALZ term with increasing magnetic field $H$. The feature that Zeman terms break down is also observed in $YBa_2Cu_3O_{7-\delta}$ (YBCO) [40]. It should be noted that this can not be explained by sample misalignment. Any angle between the $CuO_2$ planes and the applied field should give an additional orbital contribution and hence increase the magnetoconductivity. The rapid suppression seen for $\Delta\sigma_{ab}(H,T)$ with $H \perp c$ could possibly be the expected but unpredicted vanishing of superconducting fluctuations. Alternatively the theoretical expressions used for the case of $H \perp c$ are inadequate [40].

It is still a controversial issue on the treatment of the MT contributions to the magnetoconductivity in HTS. MT terms were found to disappear in the analysis of YBCO [41,42], LSCO [43], and $(Bi,Pb)_2Sr_2Ca_2Cu_3O_x$ [44] single crystals. However, some authors suggested that the MT terms significantly contribute to the magnetoconductivity [45,46]. In fact, the universal absence of the MT contributions to the magnetoconductivity was firstly observed in amorphous conventional superconductors, where strong pair-breaking effects due to thermal phonons lead to the suppression of MT terms [26]. It is worth noting that the MT terms are the only one whose presence or absence could have implications for the symmetry of the order parameter wavefunction, due to the fact that it cannot exist in the case of pure $d$-wave symmetry [47]. In the view of Yip [48] the absence of the MT terms suggests that $La_{1.6-x}Nd_{0.4}Sr_xCuO_4$ films are superconductors with $d$-wave pairing.

In the following analysis, we focus on the case of $H \parallel c$. Figure 5 shows the magnetoconductivity $\Delta\sigma_{ab}(H,T)$ with $H \parallel c$ at several temperatures for $La_{1.6-x}Nd_{0.4}Sr_xCuO_4$ films ($x = 0.1, 0.12, 0.14, 0.16$). Here, solid lines are fits using the ALO term, only with $\xi_{ab}(0)$ as free parameter. The analysis shows that $\xi_{ab}(0)$ dramatically decreases with increasing Sr content $x$. Note that $\xi_c(0)$ shows relatively small change. Therefore, the main effect of doping is simply to change the fluctuations' dimension by varying the in-plane coherence length amplitude [10].

Figure 6 shows the plots of $\xi_{ab}(0)$, $\xi_c(0)$, and anisotropy ratio $\xi_{ab}(0)/\xi_c(0)$ as a function of Sr content $x$. It is found that $\xi_{ab}(0)$ and anisotropy ratio $\xi_{ab}(0)/\xi_c(0)$ show an abrupt decrease when $x$ increases from 0.1 to 0.12. A step is then observed for $x$ = 0.12 and 0.14. It was found that the superconducting transition curve of single crystals of La$_{2-x}$Ba$_x$CuO$_4$ (LBCO) shows a fan-shape broadening in magnetic fields for $x$ = 0.08, while it shifts toward the low-temperature side in parallel with increasing field for $x$ = 0.11 and 0.12 where the charge-spin stripe order is formed at low temperatures [33]. Based on the results estimated from the analysis on fluctuation conductivity, we demonstrate that the transformation from the fan-shaped broadening to parallel shift in La$_{1.6-x}$Nd$_{0.4}$Sr$_x$CuO$_4$ films can be interpreted in terms of a dimension crossover from 2D to 3D with increasing charge carrier density, i.e. HTS-like behavior to the conventional-superconductor-like behavior. There is no need to consider magnetic-field-induced stabilization of the stripe order.

**IV. CONCLUSION**

In summary, superconducting fluctuations in La$_{1.6-x}$Nd$_{0.4}$Sr$_x$CuO$_4$ films ($x$ = 0.1, 0.12, 0.14, 0.16) are analysed by the paraconductivity and magnetoconductivity approaches. Nd doping causes a low-temperature upturn behavior of $\rho_{ab}(T)$ in La$_{1.6-x}$Nd$_{0.4}$Sr$_x$CuO$_4$ films, due to the lattice randomness effect by Nd substitution. The in-plane conductivity $\Delta\sigma_{ab}(\varepsilon)$ was extracted using a polynomial $a_1/T + a_2 + a_3T + a_4T^2$ as background resistivity. The paraconductivity are well described by the mean-field-like GGL approach extended to the high-$\varepsilon$ region by means of the introduction of a total-energy cutoff.

The magnetoconductivity, $\Delta\sigma_{ab}(H,T)$, of La$_{1.6-x}$Nd$_{0.4}$Sr$_x$CuO$_4$ films is studied in magnetic fields up to 14 T oriented both parallel and perpendicular to the $c$ axis. Good agreement is obtained for $H\parallel c$ only with ALO term. For $H\perp c$, however, the measured $\Delta\sigma_{ab}(H,T)$ can not be described with the present fluctuation theory. The negligibly small MT contributions are suggestive of anisotropic Cooper pairing, i.e., a $d$-wave pairing in La$_{1.6-x}$Nd$_{0.4}$Sr$_x$CuO$_4$ superconductors. It is found that the anisotropy of La$_{1.6-x}$Nd$_{0.4}$Sr$_x$CuO$_4$ films reduces with increasing Sr content $x$. The transformation from the fan-shaped broadening to parallel shift in La$_{1.6-x}$Nd$_{0.4}$Sr$_x$CuO$_4$ films can be interpreted in terms of a dimension crossover from 2D to 3D with increasing charge carrier density.

**Acknowledgments**

This work was supported by the National Nature Science Fund and the National Basic Research Program of China.

**Figure captions**

Fig. 1 Illustrations of the resistive transition for La$_{1.6-x}$Nd$_{0.4}$Sr$_x$CuO$_4$ film ($x$ = 0.1, 0.12, 0.14, 0.16) (open circles). The Dashed lines are the plots of the normal-state background resistivity using the polynomial $a_1/T + a_2 + a_3T + a_4T^2$. The shading region displays the discrepancy between the experimental data and the extrapolated normal-state background resistivity, which is thought to be dominated by superconducting fluctuations. The peak of $d\rho_{ab}/dT$ vs $T$ curves corresponds to the mean-field critical temperature $T_c$.

Fig. 2 The in-plane paraconductivity vs the reduced temperature $\varepsilon = \ln(T/T_c)$ for La$_{1.6-x}$Nd$_{0.4}$Sr$_x$CuO$_4$ films ($x$ = 0.1, 0.12, 0.14, 0.16). The solid lines are the best fittings to the data using Eq. (3). Dashed lines are the plots of a 2D AL formulation, i.e. Eq. (2). The dotted lines correspond to the $1/\varepsilon^3$ law.

Fig. 3 Resistive transition curves for La$_{1.6-x}$Nd$_{0.4}$Sr$_x$CuO$_4$ film ($x$ = 0.1, 0.12, 0.14, 0.16) under various applied magnetic fields for both $H \parallel c$ (left panel) and $H \perp c$ (right panel) configurations.

Fig. 4 The illustrations of the magnetoconductivity of La$_{1.6-x}$Nd$_{0.4}$Sr$_x$CuO$_4$ film for $x$ = 0.1 at $T$ = 22 K (left panel) and $x$ = 0.16 at $T$ = 35 K (right panel) as a function of $H$, respectively. Open circles and open squares correspond to the magnetoconductivity data in the case of $H \parallel c$ and $H \perp c$. The ALO, MTO, ALZ, and MTZ contributions are plotted separately. The parameters for the fits are sown in the figure.

Fig. 5 The magnetoconductivity for La$_{1.6-x}$Nd$_{0.4}$Sr$_x$CuO$_4$ films ($x$ = 0.1, 0.12, 0.14, 0.16) as a function of $H$ for $H \parallel c$ at several temperatures. The circles are experimental data and the solid lines represent theoretical results only with ALO term.

Fig. 6 The plots of in-plane coherent length $\xi_{ab}(0)$, out-of-plane coherent length $\xi_c(0)$, and anisotropy ratio $\xi_{ab}(0)/\xi_c(0)$ as a function of Sr content $x$. Solid line is a guide to the eye.

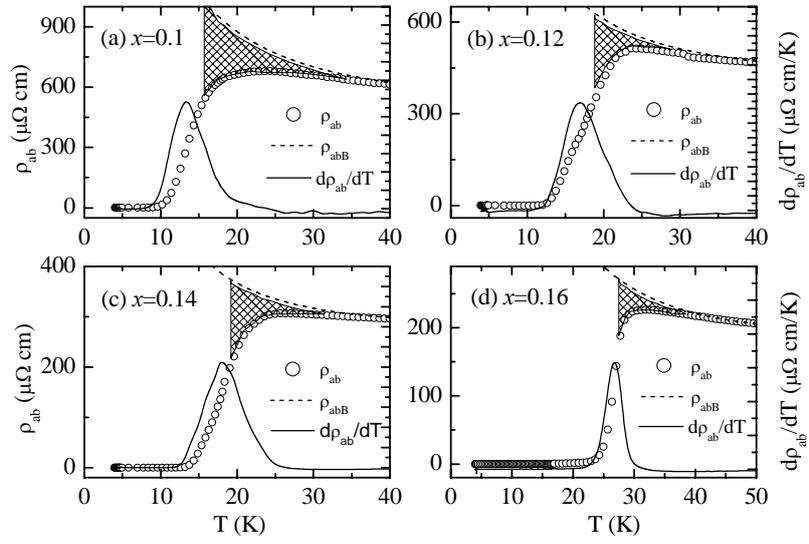

Fig. 1 by Y. Liu *et al*.

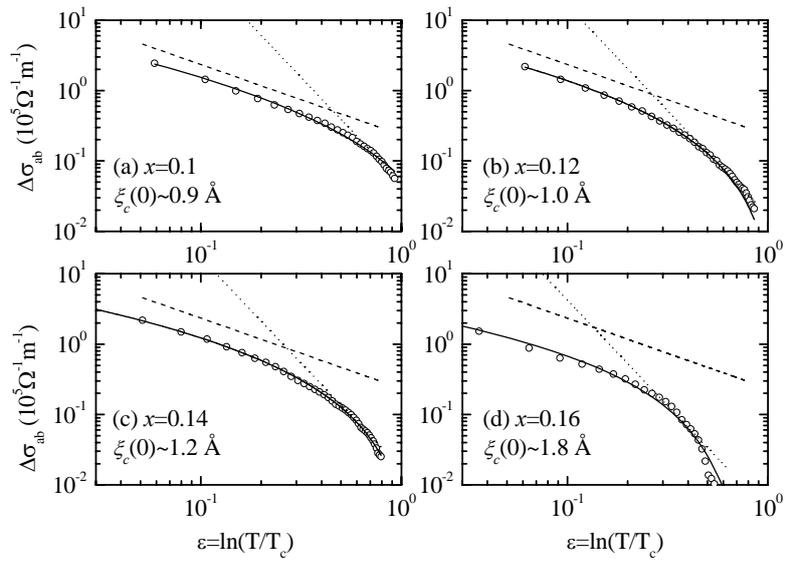

Fig. 2 by Y. Liu *et al*.

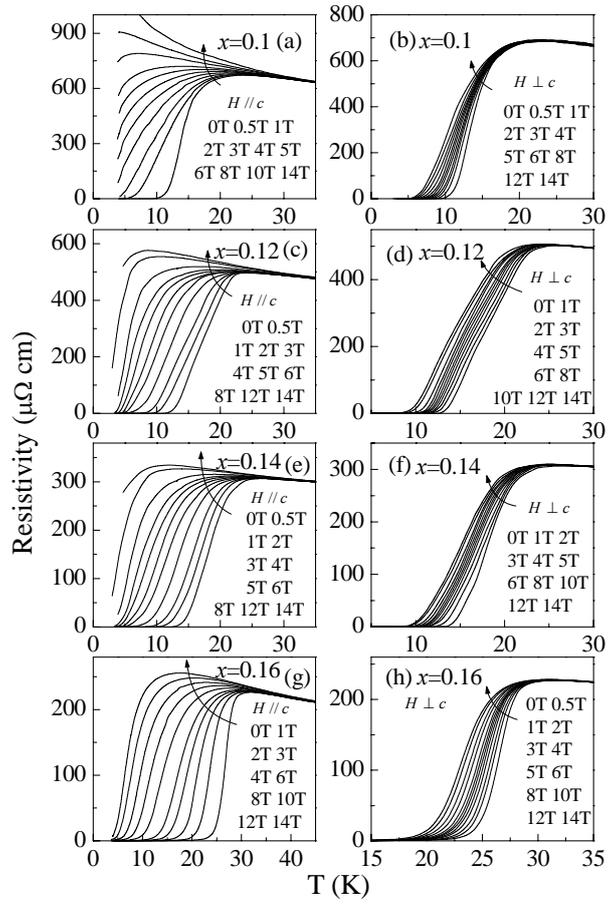

Fig. 3 by Y. Liu *et al.*

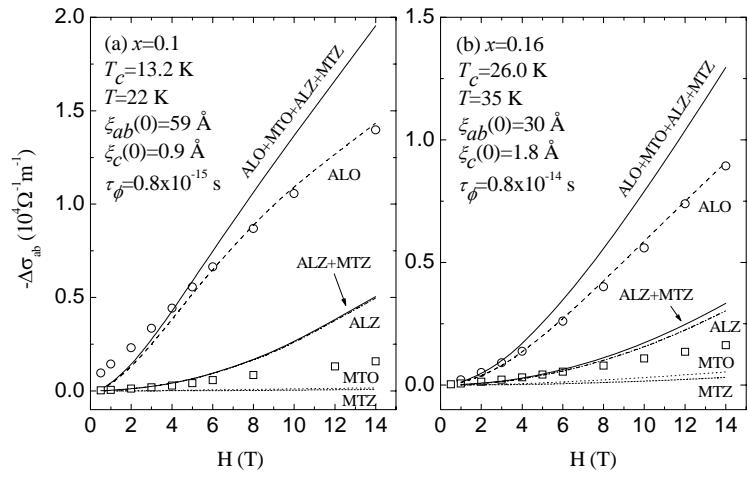

Fig. 4 by Y. Liu *et al*.

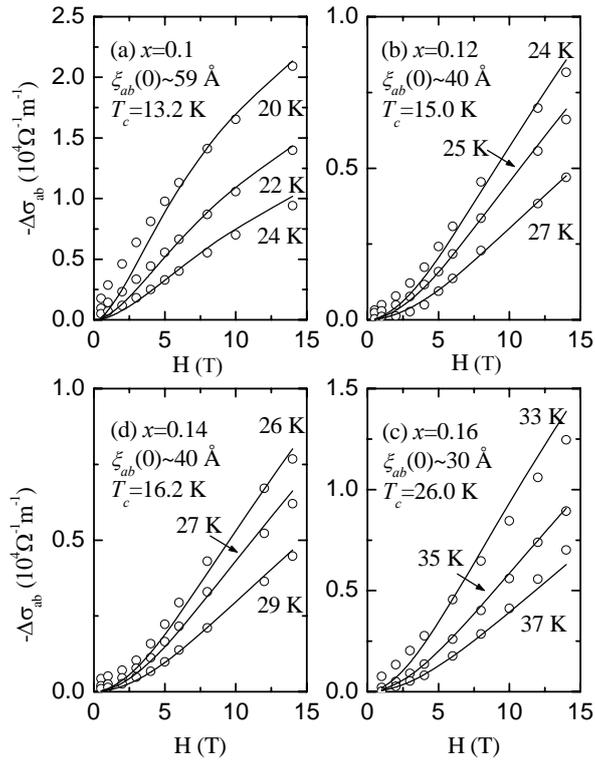

Fig. 5 by Y. Liu *et al*.

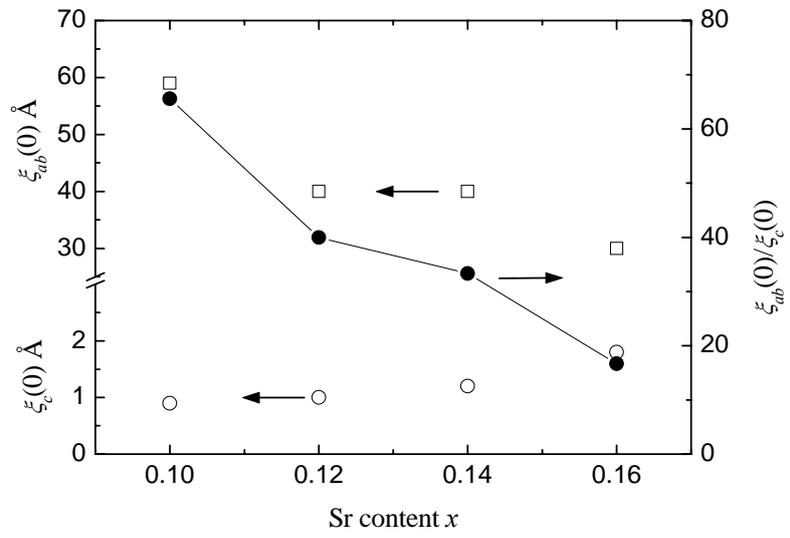

Fig. 6 by Y. Liu *et al*.